\documentclass[12pt]{iopart}
\usepackage{graphicx}
\expandafter\let\csname equation*\endcsname\relax
\expandafter\let\csname endequation*\endcsname\relax 
\usepackage{amsmath,amssymb}
\usepackage[square,comma,sort&compress,numbers]{natbib}
\begin{document}

\title[]{Current-induced cooling phenomenon 
in a two-dimensional electron gas under a magnetic field}

\author{Naomi Hirayama$^1$, Akira Endo$^2$, Kazuhiro Fujita$^2$, Yasuhiro Hasegawa$^3$, Naomichi Hatano$^4$, Hiroaki Nakamura$^5$, Ry$\bar{\rm{o}}$en Shirasaki$^6$ and Kenji Yonemitsu$^7$}

\address{
$^1$ Department of Materials Science and Technology, Faculty of Industrial Science and Technology, Tokyo University of Science, 2641 Yamazaki, Noda, Chiba 278-8510, Japan\\ 
$^2$ Institute for Solid State Physics, University of Tokyo, 5-1-5 Kashiwanoha, Kashiwa, Chiba 277-8581, Japan\\
$^3$ Department of Environmental Science and Technology, Saitama University, 255 Shimo-Okubo, Sakura, Saitama City, Saitama 338-8570, Japan\\
$^4$ Institute of Industrial Science, University of Tokyo, 4-6-1 Komaba, Meguro, Tokyo 153-8505, Japan\\
$^5$ Fundamental Physics Simulation Research Division, National Institute for Fusion Science, 322-6 Oroshi-cho, Toki, Gifu, 509-5292, Japan\\
$^6$ Department of Physics, Yokohama National University, 79-5 Tokiwadai, Hodogaya, Yokohama 240-8501, Japan\\
$^7$ Department of Physics, Chuo University, 1-13-27 Kasuga, Bunkyo, Tokyo 112-8551, Japan
}
\eads{\mailto{hirayama@rs.tus.ac.jp},\mailto{akrendo@issp.u-tokyo.ac.jp}}

\begin{abstract}
We investigate the spatial distribution of temperature induced by a dc current in a two-dimensional electron gas (2DEG) subjected to a perpendicular magnetic field.
We numerically calculate the distributions of the electrostatic potential $\phi$ and the temperature $T$ in a 2DEG enclosed in a square area surrounded by insulated-adiabatic (top and bottom) and isopotential-isothermal (left and right) boundaries (with $\phi_\mathrm{left} < \phi_\mathrm{right}$ and $T_\mathrm{left} =T_\mathrm{right}$), using a pair of nonlinear Poisson equations (for $\phi$ and $T$) that fully take into account thermoelectric and thermomagnetic phenomena, including the Hall, Nernst, Ettingshausen, and Righi-Leduc effects.
We find that, in the vicinity of the left-bottom corner, the temperature becomes lower than the fixed boundary temperature, contrary to the naive expectation that the temperature is raised by the prevalent Joule heating effect.
The cooling is attributed to the Ettingshausen effect at the bottom adiabatic boundary, which pumps up the heat away from the bottom boundary.
In order to keep the adiabatic condition, downward temperature gradient, hence the cooled area, is developed near the boundary, with the resulting thermal diffusion compensating the upward heat current due to the Ettingshausen effect. 
\end{abstract}

\maketitle

\section{\label{Introduction}Introduction}
Thermoelectric and thermomagnetic phenomena \cite{Harman2} have recently been attracting renewed interest not only as a route for potentially highly efficient device application, e.g., in refrigeration or generating electricity, but also as an effective tool to explore fundamental properties of solid state materials \cite{Zlatic,Rowe,Gallagher,Fletcher1,Goldsmid10}.
Being sensitive to the energy derivative of the electric conductivity (or the density of states) or to the entropy of the system, thermoelectric and thermomagnetic properties provide us with information on the materials complementary to, and often with higher sensitivity than, the information inferred from the electric conductivity \cite{Ying94,Chickering09,Goswami09,Yang09,Chickering,Bergman10,Behnia,Behnia3,Endo1}.
For instance, it has been shown that the Seebeck or Nernst coefficient measured in a bismuth single crystal \cite{Behnia,Behnia3} or a two-dimensional electron gas (2DEG)~\cite{Endo1} exhibits clearer quantum oscillations due to the Landau quantization compared to those of the electric conductivity (the Shubnikov-de Haas oscillations).
Moreover, thermoelectric effects introduce additional twist to the measurement of electric conductivity or resistivity, or more generally to the distribution of the electrostatic potential and electric current.
For example, it is necessary to take into consideration the thermovoltages in the precision resistivity measurement \cite{Yoshihiro}.
Further complication arises by the application of a magnetic field \cite{Cooper97,Akera1,Akera2,Komori05}.
Nontrivial distributions of the potential and current, and hence the temperature, can be generated by the thermoelectric and thermomagnetic effects.

\begin{figure*}[b]
\begin{center}
\includegraphics[width=11cm,clip]{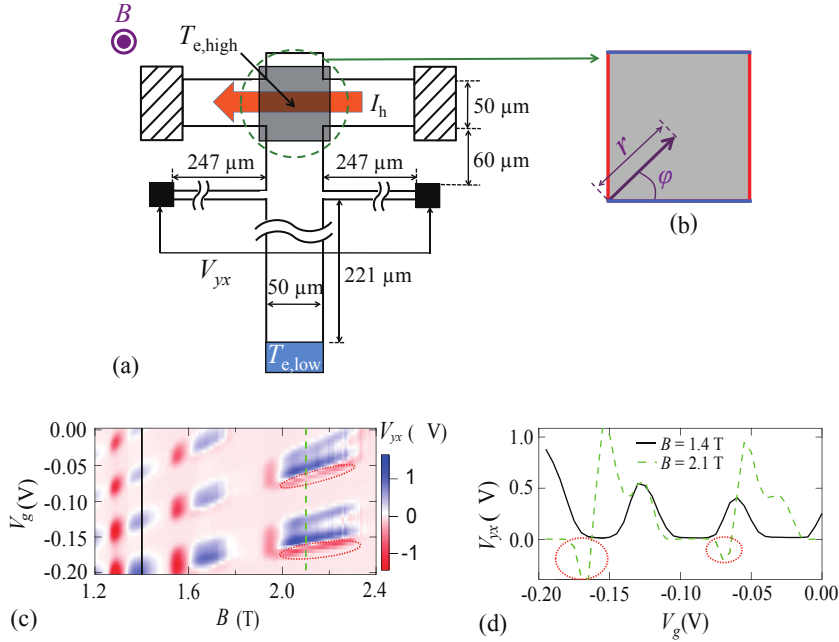} 
\caption{(a) Schematic drawing of the experimental device~\cite{Fujita}.
A magnetic field $B$ is applied perpendicular to the 2DEG plane.
The Nernst voltage $V_{yx}$ is measured between the probes depicted as the black squares.
The gray square indicates the front gate.
The current $I_\mathrm{h}=4 \, \, \mathrm{nA/\mu m}$ heats the region underneath the front gate to $T_\mathrm{e,high}$ and introduces temperature gradient toward the bottom pad held at $T_\mathrm{e,low}=40\,\,\mathrm{mK}$.
(b) Geometry used in the calculation to approximate the heater section (section beneath the front gate).
Top and bottom (blue) boundaries are assumed to be insulated and adiabatic, and left and right (red) edges have fixed potential and temperature, $\phi_{\mathrm{left}}=0\,\,\mathrm{nV}$, $\phi_{\mathrm{right}}=80\,\,\mathrm{nV}$, $T_{\mathrm{left}}=T_{\mathrm{right}}=40\,\,\mathrm{mK}$.
We introduce the polar coordinate $(r, \varphi)$ with the origin located at the left-bottom corner.
(c) Color plot of experimentally obtained Nernst voltage $V_{yx}$ in the plane of the magnetic field $B$ and the applied gate voltage $V_g$, with blue and red colors representing the positive and negative values, respectively.
(d) Cross sections indicated by solid and dashed vertical lines in (c). Anomalous behavior discussed in the text is highlighted by dotted ellipses in (c) and (d).}
\label{fig1}
\end{center}
\end{figure*}

Fujita \textit{et al.} \cite{Fujita} recently reported seemingly anomalous behavior of the Nernst signal in a quantum Hall system, which suggests possible cooling of the electron temperature by the current intended to heat the electron system and introduce the temperature gradient.
\Fref{fig1}
(a) shows a schematic diagram of the experimental device fabricated from a conventional GaAs/AlGaAs 2DEG.
The top (horizontal) bar is used as a heater; Joule heating by the heating current $I_\mathrm{h}=4\,\,\mathrm{nA / \mu m}$ raises the electron temperature $T_\mathrm{e,high}$ there and introduces temperature gradient toward the Ohmic contact pad below.
The pad is thermally connected to the mixing chamber of the dilution fridge kept at $T_\mathrm{e,low}=40\,\,\mathrm{mK}$, in which the sample is immersed.
Thermoelectric voltages are measured in the main (vertical) Hall bar.
Arms to measure the transverse thermoelectric (Nernst) voltage $V_{yx}$ are shown in the figure.
With this current heating technique, one can heat up the electron temperature selectively, leaving the lattice temperature intact (so long as $I_\mathrm{h}$ is kept low enough).
Therefore, one can pick out the diffusion contribution in the thermoelectric voltages~\cite{Maximov} and can eliminate the phonon-drag contribution, which is often the dominant contribution in the thermoelectric powers in a 2DEG~\cite{Fletcher1}. (Note, however, that the phonon-drag contribution can be negligibly small at $T=40$ mK anyway, see below).
The (negative) gate voltage $V_g $ applied to the front gate (shown by the gray square) allows us to control the carrier density and hence the resistance of the heater section independently from the main Hall bar.
\Fref{fig1}(c) shows the Nernst voltage $V_{yx}$ plotted in the $B$-$V_g$ plane.
Thermoelectric voltages vanish when the 2DEG is in the quantum Hall states~\cite{Fletcher1,JG,Oji84} and Joule heating does not work when the heater section is in the dissipationless state.
Therefore, nonvanishing signal appears only when both the main Hall bar and the heater section are in between two adjacent quantum Hall states, namely only when the Fermi energies ($E_F$) of both sections cross the (disorder-broadened) Landau levels having finite densitiy-of-states.

Below $B= 1.8 \, \, \mathrm{T}$, the Nernst voltage $V_{yx}$ behaves as expected, showing oscillations as a function of $B$ (taking negative then positive values when $E_F$ of the main Hall bar crosses a Landau level with the increase of the magnetic field)~\cite{Fletcher2,JG}, but does not depend much on $V_g$ insofar as the heater section is in the dissipative states.
Note that $V_g$ alters only the heater section and should have no effect on the main Hall bar (the section where the thermoelectric voltages are measured).
Anomalous behaviors are observed above $1.8\,\,\mathrm{T}$: $V_{yx}$ alternates the sign when $V_g$ is swept at a fixed $B$ (see \fref{fig1}(d)).
Negative $V_{yx}$ appears for smaller (more negative) $V_g$ at the higher magnetic field side where $V_{yx}$ is expected to be positive (the areas indicated by dotted ellipses in figures \ref{fig1}(c) and (d)).
The inversion of the sign of $V_{yx}$ can naively be interpreted as resulting from the inversion of the sign of the temperature gradient, implying that the ``heater" section can actually be cooled by $I_\mathrm{h}$, depending on the value of $V_g$.

This speculation led us to investigate, in the present paper, the current-induced temperature distribution of  a 2DEG placed in a magnetic field, in pursuit of the possibility of the current-induced cooling. We numerically examine the spatial distribution of the temperature $T$ in a rectangular segment of 2DEG that simulates the heater section of the experimental device, fully taking into account the thermoelectric and thermomagnetic effects.
In order to focus on the current-induced cooling, however, we consider simplest possible settings and do not make an attempt to reproduce in detail the device configuration (\fref{fig1}(a)) of the experiment delineated above; we leave out the main Hall bar altogether and approximate the heater section as a square surrounded by insulated-adiabatic (top and bottom) and isopotential-isothermal (left and right) boundaries in the calculation, as depicted in \fref{fig1}(b).
For the method of the calculation, we basically follow the prescription developed by Okumura and coauthors \cite{Okumura1} for 3D semiconductors at the room temperature and applied their treatment to a 2DEG at a low temperature.
In this treatment, the contributions of phonons are neglected, which is justified at the extremely low temperature ($40\,\,\mathrm{mK}$) considered in the present paper.
In fact, the electron-phonon scattering time $\tau_\text{e-ph}$ in a GaAs/AlGaAs heterostructure in which the 2DEG resides is estimated to be $\sim 0.01-10$ ms at $T=40$ mK \cite{Price82,Wennberg86,Mittal96}. Despite the wide variation among the literatures, $\tau_\text{e-ph}$ is still orders of magnitude larger than the elastic scattering time due to the impurity scattering (or the boundary scattering in the case of a small sample in a ballistic regime). 
The distributions of the electrostatic potential $\phi$ (with the electrochemical potential given by $-e\phi$; see the discussion below) and the temperature $T$ are obtained by solving the nonlinear Poisson equations, $\nabla^2 \phi =F(T, \nabla T, \nabla \phi)$ and $\nabla^2 T =G(T, \nabla T, \nabla \phi)$, with the functions $G$ and $F$ derived from the transport equations \cite{Okumura1,LL,Harman1,Harman2,Callen}, as will be detailed in \sref{chap2}.
We find that a magnetic field distorts equipotential lines and generates an uneven temperature distribution with high- and low-temperature areas emerging at the opposite corners of the square (see \fref{fig3}).
The low-temperature area is found to become colder than the isothermal boundaries, as briefly reported in our previous publications \cite{Hirayama1,Hirayama2}.
Two opposite corners with the temperatures lower and higher than the bath, respectively, resembling those presented here was previously also reported in numerical calculations by Ise \textit{et al.} \cite{Akera2} when the current is small enough, although the mechanism responsible for the cooling was not explicitly specified in that paper.
The emergence of the cooled part possibly give a qualitative account of the experimentally observed sign reversal in the Nernst signal \cite{Fujita} mentioned above. 
In order to clarify the relation of our numerical results to the experiment described above performed in the quantum Hall regime, however, much improvement has to be made in our theoretical treatment, as will be discussed in \sref{sec_discussion}.

The main purpose of the present paper is to clarify the mechanism of the cooling phenomenon found in our numerical calculation.
We evaluate the terms in the right hand side of the nonlinear Poisson equations $F(T, \nabla T, \nabla \phi)$, $G(T, \nabla T, \nabla \phi)$, and identify the dominant terms that induce the cooling.
We find that the cooling is mainly attributable to the adiabatic condition at the bottom edge.
It causes a temperature gradient and hence the thermal diffusion to balance out the heat current away from the edge driven by the Ettingshausen effect, and consequently generates the cooled area adjacent to the edge.
We will also show with simple analytical arguments that the cooled area, in principle, emerges by the application of an arbitrarily small magnetic field.

The paper is organized as follows.
In \sref{chap2}, we describe the method of calculating the spatial distributions of $\phi$ and $T$ by solving the nonlinear Poisson equations followed by the results of the calculations.
In \sref{chap3}, we pinpoint the mechanism of the partial cooling through the identification of the dominant terms in the equations.
In \sref{sec_discussion}, we discuss our results in connection with the experiment that motivated our study.
\Sref{chap4} is devoted to conclusions.

\section{\label{Numerical Calculation}Numerical Calculation}
\label{chap2}

The transport equations describing the electric current density $\boldsymbol{J}$ and the thermal current density $\boldsymbol{J_Q}$ for isotropic systems~\cite{Okumura1,LL,Harman1,Harman2,Callen} are:
\begin{align}
&- \nabla \phi
= \rho \boldsymbol{J}
+ R (\boldsymbol{B} \times \boldsymbol{J})
+ \alpha \nabla T 
+ N (\boldsymbol{B} \times \nabla T), 
\label{EC potential} \\
&\boldsymbol{J_Q} =
\alpha T \boldsymbol{J}  
+ N T (\boldsymbol{B} \times \boldsymbol{J}) 
- \kappa \nabla T
+ \kappa M (\boldsymbol{B} \times \nabla T), 
\label{JQ} 
\end{align}
where $\boldsymbol{B}$ denotes the magnetic field,
$\rho$ the electric resistivity,
$R$ the Hall coefficient,
$\alpha$ the Seebeck coefficient,
$N$ the Nernst coefficient,
$\kappa$ the thermal conductivity,
and $M$ the Righi-Leduc coefficient.
The transport coefficients, $\rho$, $R$, $\alpha$, $N$, $\kappa$, and $M$ are all defined in the isothermal conditions.
The terms on the right-hand sides of equations \eref{EC potential} and \eref{JQ} respectively represent transport phenomena as follows: Ohm's law, the Hall effect, the Seebeck effect, and the Nernst effect in \eref{EC potential}; the Peltier effect, the Ettingshausen effect, Fourier's law of the thermal conductivity, and the Righi-Leduc effect in \eref{JQ}.

We define the energy-flux density $\boldsymbol{J_U}$ as
\begin{align}
\boldsymbol{J_U}  = \boldsymbol{J_Q} +\phi \boldsymbol{J}.
\label{JU}
\end{align}
Here we selected the Fermi level (chemical potential at $T=0$) as the origin of the energy.
Noting that the temperature dependence of the chemical potential is negligibly small at the low temperatures considered in the present paper, the electrochemical potential is given by $-e\phi$, hence the definition \eref{JU}.
From equations \eref{EC potential}--\eref{JU} and the equations of continuity in the steady state $\nabla \cdot \boldsymbol{J} = \nabla \cdot \boldsymbol{J_U} = 0$, we obtain the nonlinear Poisson equations (see \ref{appendix1} for the derivation),
\begin{align}
\nabla^2 \phi
&=
 \rho C (T) J^2
\notag \\
&
-\left[
\frac{d \alpha}{d T}
+ \frac{R B^2}{\rho} \frac{d N}{d T} 
+ C(T)
\left(
\frac{NB^2 T}{\rho} \frac{d N}{ d T} - \frac{ d \kappa}{ d T}
\right)
\right]
( \nabla T)^2
\notag \\
&
+\left[
\frac{R}{\rho} \frac{d \rho}{d T} - \frac{d R}{d T}
-C(T) \left(
T \frac{d N}{d T} + 2N - \frac{NT}{\rho} \frac{d \rho}{d T}
\right)
\right]
 \left[
\left( \boldsymbol{B} \times \boldsymbol{J}
 \right) \cdot \nabla T 
\right]
\notag \\
&
-\left[
\frac{d \rho}{d T}
+ \frac{R B^2}{\rho} \frac{dR}{dT} 
+C(T) \left( T \frac{d \alpha}{d T} + \frac{N B^2 T}{\rho} \frac{d R}{d T}
\right)
\right]
\left(  \boldsymbol{J} \cdot \nabla T \right)
\notag \\
&
\equiv F(T, \nabla T, \nabla \phi ),
\label{PHIpoisson1}
\\
&\hspace{10pc}
C(T) 
=
\frac{\alpha \rho + R N B^2}{\rho \kappa - N^2 B^2 T},
\label{C1}
\end{align}
\begin{align}
\nabla^2 T
&=
\frac{\rho}{ \rho \kappa - N^2 B^2 T} 
\notag \\
&\times
\left\{ -\rho J^2 
+
\left(
\frac{N B^2 T}{\rho} \frac{d N}{d T}
- \frac{d \kappa}{d T}
\right)
( \nabla T)^2
 \right.
\notag \\
&+
\left(
T \frac{dN}{dT} + 2N - \frac{NT}{\rho} \frac{d \rho}{dT}
\right)
\left[
\left( \boldsymbol{B} \times \boldsymbol{J} \right) \cdot \nabla T 
\right]
\notag \\
&+ 
\left.
\left(
T \frac{d \alpha}{d T}
+\frac{N B^2 T}{\rho} \frac{d R }{d T}
\right)
\left(  \boldsymbol{J} \cdot \nabla T \right) 
\right\}
\notag \\
&\equiv G(T, \nabla T, \nabla \phi ).
\label{Tpoisson1}
\end{align}
The electric current $\boldsymbol{J}=(J_x, \, J_y)$ to be replaced in equations \eref{PHIpoisson1} and \eref{Tpoisson1} is obtained by inverting \eref{EC potential}:
\begin{align}
\begin{pmatrix}
J_x \\
J_y \\
\end{pmatrix} 
&=
\frac{1}{\rho^2 + R^2 B^2} 
\begin{pmatrix}
\rho &   R B  \\
- R B & \rho  \\
\end{pmatrix}
\notag \\
&\hspace{3pc}
\times
\begin{pmatrix}
- \partial_x \phi - \alpha \partial_x T + N B \partial_y T \\
- \partial_y \phi - \alpha \partial_y T - N B \partial_x T \\
\end{pmatrix}.
\label{J}
\end{align}
\begin{figure}[hdtp]
\begin{center}
\includegraphics[width=8.2cm,clip]{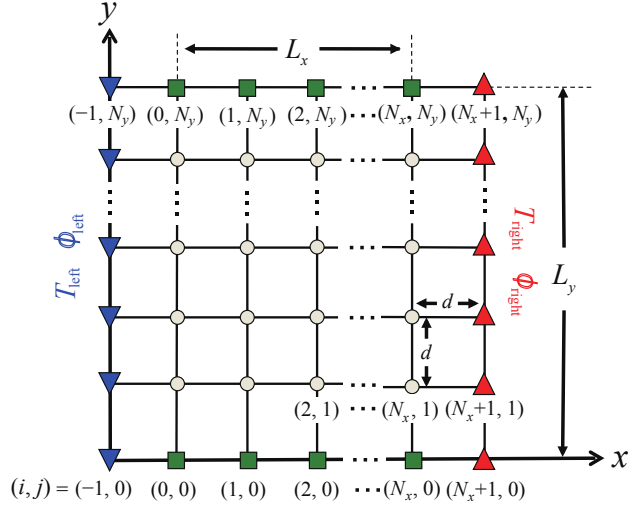} 
\caption{Discretized two-dimensional sample, where the sample size is $L_x=L_y=10\,\mathrm{\mu m}$ and the mesh size is $d=0.1\,\mathrm{\mu m}$.
Thus, the number of the grid points is $(N_x + 1) \times (N_y + 1)=101 \times 101$.
The label $(i , \, j)$ denotes the $(i, \, j)$th grid point along the $x$- and $y$-axes, respectively.
The downward (blue) and upward (red) triangles on $i=-1$ and on $i=N_x+1$ have potentials and temperatures fixed to $(\phi_{\mathrm{left}}, \, T_{\mathrm{left}})$ and $(\phi_{\mathrm{right}}, \, T_{\mathrm{right}})$, respectively, set as $\phi_\mathrm{left}=0\,\,\mathrm{nV}$ and $\phi_\mathrm{right}=80\,\,\mathrm{nV}$, $T_{\mathrm{left}} = T_{\mathrm{right}} = 40\,\,\mathrm{mK}$.
The (green) squares on the top and bottom boundaries are insulated and adiabatic.
The magnetic field $\boldsymbol{B}$ is applied perpendicular to the sample.
}
\label{fig2}
\end{center}
\end{figure}
As illustrated in \fref{fig1}(b), the left and the right edges are isopotential and isothermal, i.e., $\phi$ and $T$ are fixed (the Dirichlet conditions).
On the top and the bottom edges, we set insulated and adiabatic conditions; the normal components of $\boldsymbol{J}$ and $\boldsymbol{J_Q}$ vanish at these boundaries.
Let us denote the quantities at the top and bottom boundaries with tilde, $\widetilde{\phi}$, $\widetilde{\boldsymbol{J}}$, $\widetilde{T}$, and $\widetilde{\boldsymbol{J}}_{\boldsymbol{Q}}$.
Equations \eref{EC potential} and \eref{JQ} at these boundaries are then reduced to
\begin{subequations}
\label{eqsBC}
\begin{align}
- \partial_x \widetilde{\phi}
&= \rho \widetilde{J}_x + \alpha \partial_x \widetilde{T} - N B \partial_y \widetilde{T},
\label{APB1}
\\
- \partial_y \widetilde{\phi}
&=  R B \widetilde{J}_x  +  \alpha \partial_y \widetilde{T}  + N B \partial_x \widetilde{T},
\label{APB2} \\
0=&
N T B \widetilde{J}_x -\kappa \partial_y \widetilde{T} + \kappa MB \partial_x \widetilde{T}.
\label{eqsBCT}
\end{align}
\end{subequations}
We thus obtain the following derivatives (the Neumann conditions),
\begin{subequations}
\label{BC}
\begin{align}
\partial_y \widetilde{\phi}
=&
- ( R  + \frac{\alpha N T}{\kappa} ) 
B \widetilde{J}_x 
- (\alpha M +N) B \partial_x \widetilde{T},
\label{BCPHI1} \\
 \partial_y \widetilde{T} =&
\frac{N T B}{\kappa} \widetilde{J}_x  +  MB \partial_x \widetilde{T},
\label{BCT1}
\end{align}
where the electric current $\widetilde{J}_x$ on the boundary is given by
\begin{align}
\widetilde{J}_x   &=
- \frac{ \kappa}{ \rho \kappa - N^2 B^2 T }
\left[
\partial_x  \widetilde{\phi} +( \alpha -  M N B^2) \partial_x \widetilde{T} 
\right].
\label{BCJ1}
\end{align}
\end{subequations}

We numerically solved the set of equations \eref{PHIpoisson1}--\eref{Tpoisson1} self-consistently on a discretized sample illustrated in \fref{fig2} with the Dirichlet and Neumann conditions mentioned above.
We employed the successive over-relaxation (SOR) method~\cite{FDM,Numericalrecipe} in the finite-difference calculations. 
%
We used the following strategy to achieve efficient convergence in the calculation.
We started our calculation without a magnetic field, giving the initial distributions of $\phi$ and $T$ as linear functions between the left and right isopotential-isothermal boundaries. 
Once we reached the convergence, we then increased $B$ from $0$ to $10^{-3}\,\,\mathrm{T}$ and sought the convergence.
We successively increased $B$ step by step with an increment of $10^{-3}\,\,\mathrm{T}$, using the result for the previous $B$ as the initial distribution for the next $B$.
Nevertheless, we have so far reached only up to $0.2\,\,\mathrm{T}$, the convergence becoming increasingly slower with increasing magnetic field.

We performed the calculations with the parameters listed in \tref{table1}: $\rho$ and $R$ are taken from the experimental data \cite{Fujita} and other parameters, $\alpha$, $N$, $\kappa$, $M$, and their temperature derivatives are calculated by substituting the values of $\rho$ and $R$ into the semiclassical formulas \cite{Fletcher1} (for $\alpha$ and $N$) and making use of the Wiedemann-Franz law \cite{ZimanEP} (for $\kappa$ and $M$) valid for the diffusion contribution.
\begin{table*}
\caption{\label{table1} Parameters used for the calculation.}
\begin{indented}
\item[]\begin{tabular}{cccc}
\br
Parameter & Value & Parameter & Value \\
\mr
Resistivity $\rho$
& $20.0 \, \, \mathrm{\Omega}^\mathrm{a}$
 &  $d\rho / dT$ & $1\, \,\mathrm{n \Omega K^{-1}}^\mathrm{c}$
 \\
Hall coefficient $R$ & $-1600.0 \, \, \mathrm{\Omega T^{-1}}^\mathrm{a}$ & $dR / dT$ & $1\,\,\mathrm{n \Omega T^{-1} K^{-1}}^\mathrm{c}$ \\
Seebeck coefficient $\alpha$ & $-0.175\, \, \mathrm{\mu VK^{-1}}^\mathrm{b}$& $d\alpha / dT$ & $-4.38\, \, \mathrm{\mu V K^{-2}}^\mathrm{b}$\\
Nernst coefficient $N$ & $-7.35\, \, \mathrm{\mu V K^{-1} T^{-1}}^\mathrm{b}$ & $dN / dT$ & $-184\,\,\mathrm{\mu V K^{-2} T^{-1} }^\mathrm{b}$ \\ 
Thermal conductivity $\kappa$ & $42.7\, \, \mathrm{p W K^{-1}}^\mathrm{b}$ & $d\kappa / dT$&$1\, \, \mathrm{n W K^{-2}}^\mathrm{b}$ \\
Righi-Leduc coefficient $M$ & $-70.0 \, \, \mathrm{m^2 V^{-1} s^{-1}}^\mathrm{b}$  &  $dM/dT$ & $0^\mathrm{b}$ \\
\br
\end{tabular}
\item[]{$^a$ Experimental parameters taken from \cite{Fujita}.}
\item[]{$^b$ Values calculated with semiclassical theories at $T= 40\, \, \mathrm{mK}$ using experimentally obtained $\rho$ and $R$, assuming that the scattering time $\tau$ depends on the energy $\varepsilon$ as $\tau \propto \varepsilon^{1.5}$.}
\item[]{$^c$ Tentative values for the numerical calculation, assumed to be small enough not to affect the result of the calculation.}
\end{indented}
\end{table*}
As an initial step, we used the values of the parameters at $B=0$ throughout the calculation in the present study, neglecting their magnetic-field dependence.
Although this appears to be very crude approximation, the parameters at $B=0$ represent roughly the right order of magnitude for their values under finite $B$ when the Fermi energy $E_F$ crosses the Landau levels (by contrast,  some of the parameters vanish in the quantum Hall states), and we believe they suffice to discuss qualitatively what happens when the heater section is in the dissipative regime.

\begin{figure*}[b]
\includegraphics[width=16cm,clip]{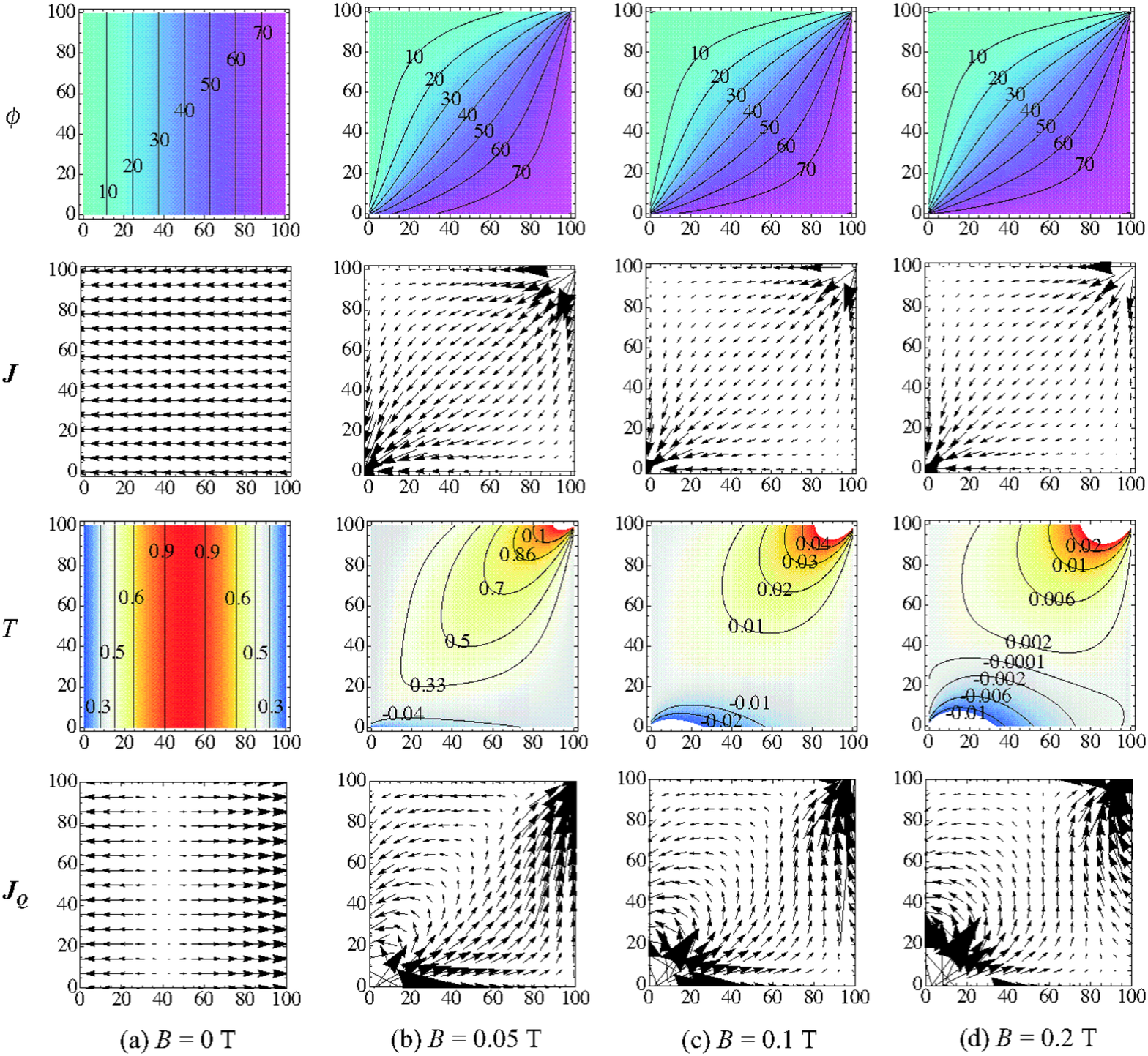}
\caption{Distributions of $\phi$, $\boldsymbol{J}$, $T$ and $\boldsymbol{J_Q}$ in $(i,j)$ plane obtained from the nonlinear Poisson equations \eref{PHIpoisson1} and \eref{Tpoisson1} with the boundary conditions $\phi_{\mathrm{left}}=0.0 \, \, \mathrm{n V}$, $\phi_{\mathrm{right}}=80\, \, \mathrm{n V}$ and $T_{\mathrm{left}}=T_{\mathrm{right}}=40\, \, \mathrm{mK}$.
The $\phi$-contours are labeled by the values in the unit of nV, and the $T$-contours by the difference from $T_{\mathrm{left}}=T_{\mathrm{right}}$ in the unit of $\mu$K\@.
(a) $B=0 \, \, \mathrm{T}$. The potential gradient generates an uniform current distribution and the resulting Joule heat raises the temperature in the mid part.
(b)--(d) $B > 0$ T\@. The magnetic field causes the distortion of the equipotential lines and the distribution of the electric current $\boldsymbol{J}$.
It also affects the distribution of the temperature $T$ and generates the area having the temperature lower than that of the heat baths.
The cooled area occurs, in principle, with an arbitrarily small magnetic field, although it is apparent above $0.03\,\,\mathrm{T}$ in our discretized sample.
(See \sref{Minimal} for discussion).}
\label{fig3}
\end{figure*}

\begin{figure*}[b]
\includegraphics[width=16cm,clip]{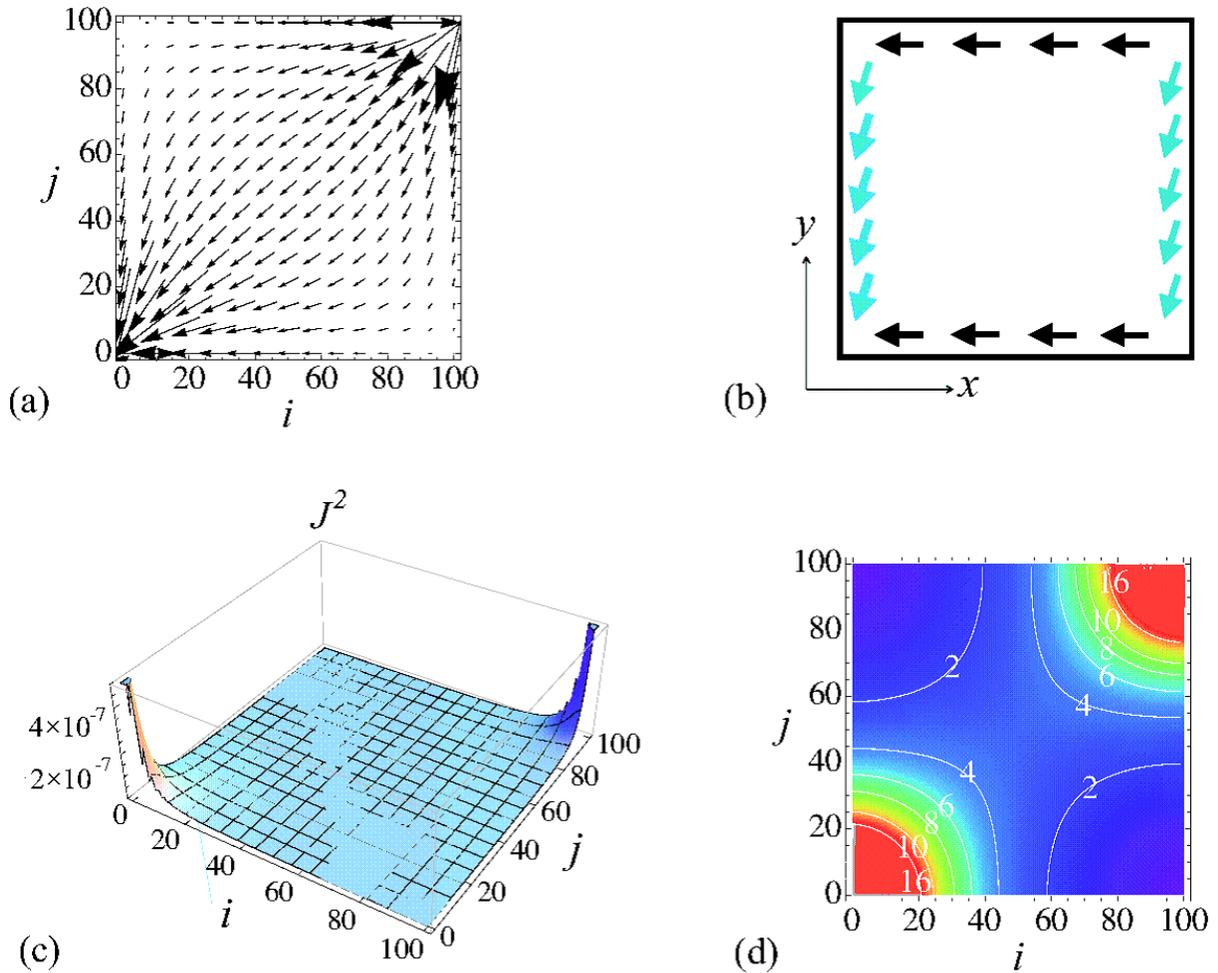}
\caption{(a) Distribution of $\boldsymbol{J}$. (b) Schematic view of $\boldsymbol{J}$ on the boundaries.
(c) Distribution of $J^2$.
(d) Contour lines of $J^2$ in units of (nA)$^2$. All data are for $B=0.1\,\,\mathrm{T}$.}
\label{fig4}
\end{figure*}

\Fref{fig3}
shows the distributions of the electrostatic potential $\phi$, the electric current $\boldsymbol{J}$, the temperature $T$, and the heat current $\boldsymbol{J_Q}$ in the case $\phi_{\mathrm{left}} =0\,\,\mathrm{nV}$, $\phi_{\mathrm{right}}=80\,\,\mathrm{nV} $, and $T_{\mathrm{left}}=T_{\mathrm{right}}=40\,\,\mathrm{mK}$.
We briefly reported the numerical results shown here in previous articles~\cite{Hirayama1, Hirayama2}. 
At $B=0$ (\fref{fig3}(a)) the electric current $\boldsymbol{J}$ flows homogeneously perpendicular to the $\phi$-contours.
The temperature, raised by the Joule heating, has a symmetric distribution decreasing towards the left and right edges held at the fixed temperature.
The asymmetry of the heat current $\boldsymbol{J}_Q$, with the net flow going to the right, arises because of the left-going potential energy flow due to $\boldsymbol{J}$ (the second term in \eref{JU}).
Once the magnetic field is switched on (figures \ref{fig3}(b)-(d)), the $\phi$-contour is distorted.
The current $\boldsymbol{J}$ flows nearly parallel to the $\phi$-contours, or more precisely, approximately at the Hall angle
\begin{align}
\theta_{\mathrm{H}} = - \arctan \left(\frac{RB}{\rho} \right),
\label{Hall}
\end{align}
deflected from the gradient $\nabla \phi$. The Hall angle $\theta_\mathrm{H}$ equals $76^\circ$, $83^\circ$, and $86^\circ$ at $B=0.05\,\,\mathrm{T}$, $0.1\,\,\mathrm{T}$, and $0.2\,\,\mathrm{T}$, respectively. 
The current $\boldsymbol{J}$ is highly concentrated at the right-top and left-bottom corners (see also \fref{fig4}).
The distributions of $\phi$ and $\boldsymbol{J}$ are qualitatively the same as well-known distributions calculated without taking the thermoelectric and thermomagnetic effects into consideration~\cite{Wick54,Rendell,Wakabayashi,Neudecker87}.
More quantitative comparison will be made in \sref{simple}.
The temperature distribution becomes asymmetric, with high and low-temperature areas emerging around the right-top and left-bottom corners, respectively (where $\boldsymbol{J}$ has a high concentration).
The low-temperature part has temperatures lower than the temperature $T_\mathrm{left}=T_{\mathrm{right}}$ of the heat baths.
We will show in \sref{Minimal} that, in principle, the cooling appears with an arbitrarily small magnetic field.

\section{Mechanism of the cooling phenomenon}
\label{chap3}
In the present section, we investigate the origin of the cooling phenomenon.

\subsection{Simplification of the governing equations}
\label{simple}

As an initial step toward the understanding of the cooling mechanism, we deduce an approximate version of the governing equations and the boundary conditions much simpler than the original ones.
To this end, we compare the terms in the relevant equations using the numerical solutions presented in the previous section, and eliminate the terms whose contributions are negligibly small compared to the other terms.
First, in equation \eref{EC potential}, we find that $|\alpha \nabla T| / |\nabla \phi| \lesssim 10^{-5}$ and $|N (\boldsymbol{B} \times \nabla T)| / |\nabla \phi| \lesssim 10^{-6}$ in the magnetic field range ($0 - 0.2$ T) examined in the present paper.
Therefore the terms $\alpha \nabla T$ and $N (\boldsymbol{B} \times \nabla T)$ can safely be neglected.
Noting that $d\rho/dT$ and $dR/dT$ are also negligibly small in a 2DEG at $T<0.1\,\,\mathrm{K}$, we arrive at the Laplace equation,
\begin{align}
\nabla^2 \phi
&=
0,
\label{simplePHI}
\end{align}
for the electrostatic potential, and the expression
\begin{align}
&\begin{pmatrix}
J_x \\
J_y \\
\end{pmatrix}     
=
 \frac{1}{\rho^2 + R^2 B^2} 
\begin{pmatrix}
\rho &   R B  \\
- R B & \rho  \\
\end{pmatrix}
\begin{pmatrix}
-\partial_x \phi  \\
-\partial_y \phi  \\
\end{pmatrix}.
\label{simpleJ}
\end{align}
for the current, as simplified equations to take the place of equations \eref{PHIpoisson1} and \eref{J}, respectively.
We can thus calculate with high accuracy the distributions of $\phi$ and $\boldsymbol{J}$ neglecting the thermoelectric and thermomagnetic effects.
This is to be expected since we adopted rather large potential difference but no difference in the temperature between left and right edges as the boundary condition.
Since the Laplace equation \eref{simplePHI} can be solved analytical for our boundary conditions \cite{Rendell} (see \sref{Minimal}), this approximation vastly simplifies the calculation.  

Next we examine equation \eref{Tpoisson1}.
We find that the term including $\rho J^2$ (corresponding to the Joule heating) is by far the dominant term, exceeding the other terms by factor $10^6$.
Along with $N^2 B^2 T / \rho \kappa \sim O (10^{-16})$, we have 
\begin{align}
\nabla^2 T
&=
-\frac{\rho}{ \kappa } J^2,
\label{simpleT} 
\end{align}
as a simplified approximate nonlinear Poisson equation for the temperature.

The boundary conditions \eref{APB1}--\eref{eqsBCT} at the top and bottom boundaries can also be simplified as,
\begin{subequations}
\label{simpleeqsBC}
\begin{align}
- \partial_x \widetilde{\phi}
&= \rho \widetilde{J}_x,
\label{simpleeqsBCPHI1}
\\
- \partial_y \widetilde{\phi}
&=  R B \widetilde{J}_x,
\label{simpleeqsBCPHI2}
\\
0=&
N T B \widetilde{J}_x -\kappa \partial_y \widetilde{T} + \kappa MB \partial_x \widetilde{T},
\label{simpleeqsBCT}
\end{align}
\end{subequations}
(Note that \eref{simpleeqsBCT} is the same as \eref{eqsBCT} because all the terms are of comparable order of magnitude and therefore cannot be neglected.)
Or equivalently, we have the Neumann condition
\begin{subequations}
\label{simpleBC}
\begin{align}
\partial_y \widetilde{ \phi }  &= - RB \widetilde{J}_x,
\label{simpleBCPHI}
\\
 \partial_y \widetilde{T} &=  \frac{NTB}{\kappa} \widetilde{J}_x +MB \partial_x \widetilde{T},
\label{simpleBCT}
\\
\widetilde{J}_x  &= - \frac{1}{\rho} \partial_x \widetilde{\phi}.
\label{simpleBCJ}
\end{align}
\end{subequations}
%
Equation \eref{simpleT} reveals that the spatial variation of the temperature in the interior of the sample is mainly determined by the Joule heating, with the thermoelectric and thermomagnetic effects playing only minor roles.
This is not the case at the boundaries, where the adiabacity is achieved among the Ettingshausen effect, thermal diffusion, and the Righi-Leduc effect, as seen in  \eref{simpleeqsBCT}.

To confirm the appropriateness of the simplification, we calculated $\phi$, $\boldsymbol{J}$, $T$, $\boldsymbol{J_Q}$ with the simplified equations \eref{simplePHI} --\eref{simpleT} and \eref{simpleBC} and obtained the distribution virtually indistinguishable from \fref{fig3}.
The relative differences $\delta A = | (A - A') / A|$, where $A$ and $A'$ are the values from the original and the simplified equations, respectively, were sufficiently small for $\phi$, $T$, $\boldsymbol{J}$, and $\boldsymbol{J_Q}$ as $\delta \phi \lesssim O (10^{-5})$, $\delta T \lesssim O (10^{-10})$, $\delta J_x \lesssim O (10^{-6})$, $\delta J_y \lesssim O (10^{-6})$, $\delta J_{Qx} \lesssim O (10^{-7})$, and $\delta   J_{Qy} \lesssim O (10^{-7})$.
It verifies that the simplified equations \eref{simplePHI}--\eref{simpleT} and \eref{simpleBC} effectively give the same distributions as those from the original equations \eref{PHIpoisson1}--\eref{J} and \eref{BC}.

\subsection{The role of the Ettingshausen and the Righi-Leduc effects}
\label{cooling mechanism}

In this section, we examine the role played by the first and the third terms in \eref{simpleeqsBCT} (the Ettingshausen and Righi-Leduc effects).
We substitute $0$ into the coefficients $N$ and/or $M$, and see whether cooled area appears or not.
We confirm that it is the Ettingshausen effect at the boundary that is indispensable for the cooling effect.

\begin{figure*}[b]
\includegraphics[width=16cm,clip]{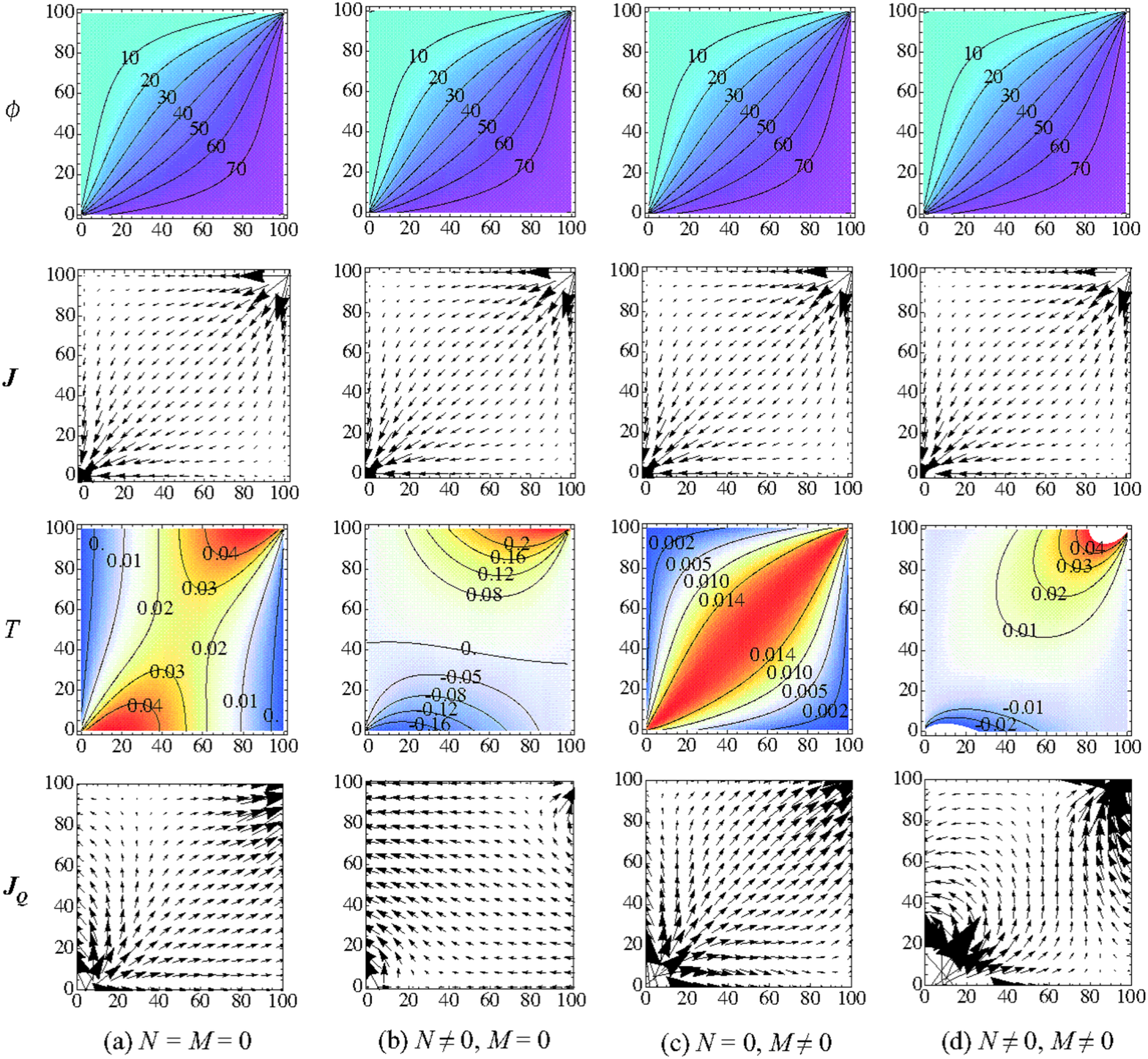}
\caption{Distributions of $\phi$, $\boldsymbol{J}$, $T$ and $\boldsymbol{J_Q}$ in $(i,j)$ plane at $B=0.1\,\,\mathrm{T}$ for (a) $N=M=0$, (b) $N \ne 0, \,\, M=0$, (c) $N = 0, \,\, M \ne 0$ and (d) $N \ne 0, \,\, M \ne 0$. (The data in (d) is the reproduction of \fref{fig4}(c).)
The $\phi$-contours are labeled in the unit of nV\@.
The labels for the $T$-contours indicate the differences from $T_{\mathrm{left}}=T_{\mathrm{right}}=40\,\,\mathrm{mK}$ in the unit of $\mathrm{\mu K}$.}
\label{fig5}
\end{figure*}

First, we consider the case $N=M=0$ in \eref{simpleeqsBCT}; that is, $\partial_y \widetilde{T}=0$ at the top and the bottom boundaries.
\Fref{fig5}(a)
shows the distributions of $\phi$, $\boldsymbol{J}$, $T$ and $\boldsymbol{J_Q}$ in this case.
Here and in what follows, we used a magnetic field $B=0.1$ T as a typical example.
The result shows that the right-top and the left-bottom corners have much higher temperatures than elsewhere owing to the Joule heating by the highly concentrated electric-current density $\boldsymbol{J}$ illustrated in \fref{fig4}.

Next, we assume that $N \ne 0$ and $ M=0$ in \eref{simpleeqsBCT}, namely, $0= NTB  \widetilde{J_x} -\kappa \partial_y \widetilde{T}$.
In this case, we obtain a cooled area, as shown in \fref{fig5}(b).
It is obvious, therefore, that the first term in \eref{simpleeqsBCT} (the Ettingshausen effect) plays a major role in the cooling.
The temperatures on the top (bottom) boundary are now higher (lower) compared with those in \fref{fig5}(a).
This change in the temperature map results from an upward temperature gradient $\partial_y \widetilde{T} = (NTB/\kappa) \widetilde{J}_x$ on the adiabatic boundaries (with the negative value of $N$).
\begin{figure*}
\begin{center}
\includegraphics[width=16cm,clip]{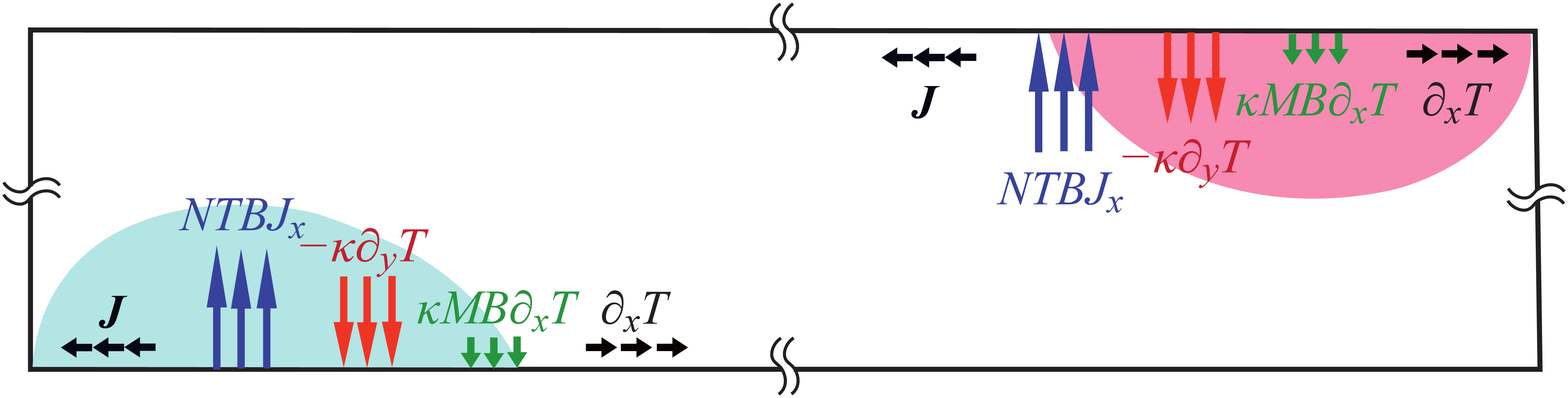} 
\caption{Schematic diagram showing the thermal current flow around the left-bottom and right-top corners.
On the top and bottom boundaries, the total heat current, which is the sum of the heat current by the thermal diffusion ($-\kappa \partial_y \widetilde{T}$, red), the Ettingshausen ($NTB\widetilde{J}_x$, blue), and the Righi-Leduc ($\kappa M B \partial_x \widetilde{T}$, green) effects, vanishes.
The heat currents by the Ettingshausen and Righi-Leduc effects go upward and downward, respectively, since $\widetilde{J}_x < 0$, $\partial_x \widetilde{T} > 0$, $N<0$, and $M<0$.
The right-top and left-bottom corners have temperatures higher and lower than the right and left boundaries, respectively, since $\partial_y \widetilde{T}>0$.
}
\label{fig6}
\end{center}
\end{figure*}
The temperature gradient can be viewed as being generated so that the resulting downward thermal diffusion $-\kappa \partial_y \widetilde{T}$ cancels the upward heat current brought about by the Ettingshausen effect $NTB\widetilde{J}_x$ at the adiabatic boundary, as illustrated in \fref{fig6}.
If we set, in turn, $N=0, \,\, M \ne 0$ in \eref{simpleeqsBCT}, i.e., $0= -\partial_y \widetilde{T} + M B \partial_x \widetilde{T}$, the cooling effect does not appear as seen in \fref{fig5}(c).
The boundary condition yields a downward gradient $\partial_y \widetilde{T} =  M B \partial_x \widetilde{T}$, with $M<0$ and $\partial_x T>0$ at the boundaries.

Finally, with $N \ne 0$ and $M \ne 0$ in \eref{simpleeqsBCT}, we obtain the distributions shown in \fref{fig5}(d), which is the same as \fref{fig3}(c) but re-presented for comparison.
Since the term $M B \partial_x \widetilde{T}$ (the Righi-Leduc effect) reduces $\partial_y \widetilde{T}$, a cooled area becomes smaller than that for $N \ne 0, \,\, M = 0$ (\fref{fig5}(b)). 

The distributions of $\phi$ and $\boldsymbol{J}$ remain virtually unaltered throughout figures \ref{fig5}(a)--(d), despite the change in the boundary condition \eref{simpleeqsBCT}.
This is because they are basically decoupled from the thermoelectric and thermomagnetic effects in the present situation, as demonstrated in \sref{simple}.

\subsection{Threshold magnetic field for the cooling}
\label{Minimal}

In this section, we show that a cooled area is generated, in principle, by an arbitrarily small magnetic field; we can always find a cooled area if we can approach indefinitely close to the left-bottom corner.
In practice, however, the minimum distance from the corner is limited by a certain physical length scale (obviously, the length, e.g., much smaller than the inter-atomic distance of the host crystal does not make sense), which sets a threshold for the magnetic field to generate the cooled area.
In our discretized system used for the numerical calculation, the minimum distance is the separation between the grid points (not the physical length scale but rather an artificial distance), which determines the lowest magnetic field for the cooled areas to be observed on our grid points.

As described in \sref{simple}, equations \eref{PHIpoisson1}--\eref{J} and \eref{BC} are well-approximated by the simplified equations \eref{simplePHI}--\eref{simpleT} and \eref{simpleBC} for the present system.
Equation \eref{simplePHI} under the boundary conditions \eref{simpleBCPHI} and \eref{simpleBCJ} can be solved analytically \cite{Rendell}. Although we have limited ourselves to a square sample thus far, analytical solutions are given more generally for rectangular samples ($L_x \ne L_y$ in \fref{fig2}). The electric field $\boldsymbol{E}=- \nabla \phi$ is given by
\begin{subequations}
\label{ExEy}
\begin{align}
E_x &= -E_0 e^{\gamma} \cos \vartheta, \label{Ex} \\
E_y &= E_0 e^{\gamma} \sin \vartheta,
\label{Ey}
\end{align}
\end{subequations}
with
\begin{align}
\gamma = - 4 \theta_{\mathrm{H}} \sum_{n=1}^{\infty}
&\frac{1}{(2n-1) \pi}
\frac{
\sinh 
\left[ (2n-1) \pi \eta  \right] 
\cos \left[ (2n-1) \pi \xi \right] 
}{
\cosh \left[ (2n-1) \frac{\pi \lambda}{2} \right]
},
\label{gamma}
\end{align}
\begin{align}
\vartheta = 4  \theta_{\mathrm{H}} \sum_{n=1}^{\infty}
&\frac{1}{(2n-1) \pi}
\frac{
\cosh 
\left[ (2n-1) \pi  \eta \right] 
\sin \left[ (2n-1) \pi \xi \right]
 }{
\cosh \left[ (2n-1) \frac{\lambda \pi }{2} \right]
 },
\end{align}
where we introduced normalized coordinates $\xi = x / L_x$, $\eta = (y-L_y/2)/ L_x$ and the aspect ratio $\lambda = L_y / L_x$.
The constant $E_0$ is determined by the potential difference between the side boundaries $\Delta \phi=\phi_{\mathrm{right}} - \phi_{\mathrm{left}}$, the Hall angle $\theta_{\mathrm{H}}$ in \eref{Hall} (or the magnetic field), and the aspect ratio $\lambda$ as
\begin{align}
&E_0 
= 
\frac{\Delta \phi}{I (\theta_{\mathrm{H}}, \alpha) L_x},
\label{Eq:E0} \\
&I  (\theta_{\mathrm{H}}, \lambda)
\equiv
\int_0^1
\cos \left\{
4 \theta_{\mathrm{H}}
\sum_{n=1}^{\infty}
\frac{\sin \left[ (2n-1) \pi \xi \right]}{(2n-1) \pi}
\mathrm{sech} 
\left[ (2n-1) \frac{\pi}{2} \lambda
\right]
\right\}
d \xi
\label{Eq:I}
\\
&\hspace{3.5pc}
\simeq J_0 ( \frac{4\theta_{\mathrm{H}}}{\pi}
\mathrm{sech} \frac{\lambda \pi}{2} ),
\end{align}
with $J_0(x)$ the Bessel function of order zero.

From equations \eref{simpleJ} and \eref{ExEy}, we have the electric current density,
\begin{subequations}
\begin{align}
J_x &= -E_0  \frac{\cos  \theta_{\mathrm{H}}}{\rho} e^{\gamma} \cos (\vartheta -  \theta_{\mathrm{H}}) , \\
J_y &= E_0  \frac{\cos  \theta_{\mathrm{H}}}{\rho} e^{\gamma} \sin (\vartheta -  \theta_{\mathrm{H}}),  \\
J&=\sqrt{J_x^2+J_y^2}=E_0 e^{\gamma} \frac{\cos  \theta_{\mathrm{H}}}{\rho}.
\label{EQ23}
\end{align}
\end{subequations}
We obtain the total current $J_\mathrm{tot}$ by integrating $J_x$ along an arbitrary axis in $y$-direction,
\begin{align}
J_\mathrm{tot}
&= - L_x \int_{-\lambda /2}^{\lambda /2} J_x \left( \xi= \mathrm{const.} ; \eta \right) d \eta
\notag \\
&= L_x E_0 \frac{ \cos (\theta_{\mathrm{H}}) }{\rho} K (\theta_{\mathrm{H}}, \lambda ),
\label{Jtot}
\end{align}
\begin{align}
&K (\theta_{\mathrm{H}}, \lambda ) 
\equiv
\lambda
\int_0^1 \cos 
\left(
4 \theta_{\mathrm{H}} \sum_{n=1}^\infty
(-1)^{n-1}
\right.
\notag \\
&\hspace{2pc}
\left.
\frac{\mathrm{cosh} 
\left[ (2n-1) \frac{\lambda \pi}{2} \eta^\prime \right]
}{(2n-1) \pi}
\mathrm{sech}
\left[
(2n-1) \frac{\lambda \pi}{2}
\right]
- \theta_{\mathrm{H}}
\right)
d \eta^\prime.
\end{align}
For $\lambda = 1$ (square sample), it can be shown, to an extremely good approximation, that $K (\theta_{\mathrm{H}}, 1) $ $\simeq$ $I (\theta_{\mathrm{H}}, 1) $, which monotonically decreases with increasing $\theta_\mathrm{H}$ from 1 at $\theta_{\mathrm{H}} = 0$ to $0.847$ at $\theta_{\mathrm{H}} =\pi / 2$. Therefore we have
\begin{align}
J_\mathrm{tot}(\lambda = 1) \simeq \frac{\Delta \phi \cos (\theta_\mathrm{H})}{\rho}.
\label{JtotDphi}
\end{align}

\begin{figure*}[b]
\includegraphics[width=16cm,clip]{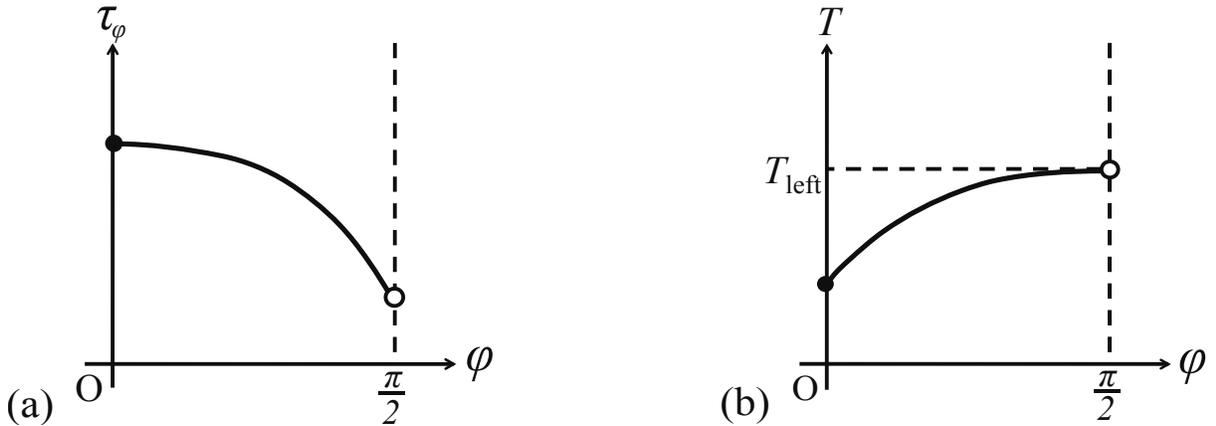}
\caption{Schematic illustration of the $\varphi$-dependences of (a) $\tau_\varphi = r^{-1} \partial_{\varphi} T$ and (b) $T$ around the left-bottom corner.
The points at $\varphi = 0$ (solid circle) and $\varphi = \pi / 2$ (open circle) correspond to the bottom and the left edges, repectively.
$\tau_\varphi (\varphi=0) > 0$ and $T (\varphi=\pi/2) = T_\mathrm{left}$ have fixed values determined by the boundary conditions.
$\tau_\varphi$ decreases with increasing $\varphi$, following \eref{eq25}.
If $\tau_\varphi (\varphi = \pi/2) > 0$, $T$ increases with increasing $\varphi$ near the left edge, ensuring the presence of the area colder than $T_\mathrm{left}$.
}
\label{fig7}
\end{figure*}

We introduce here the polar coordinate $(r, \varphi)$, where $r=\sqrt{x^2 + y^2}$ and $\varphi = \tan^{-1} (y/x)$, with the origin located at the left-bottom corner of the system as shown in \fref{fig1}(b), noting that $J$ is nearly isotropic in the vicinity of the corner as can be seen in \fref{fig4}(c) and (d).
Using the polar coordinate notation for the temperature gradient, $\tau_r \equiv \partial_r T$ and $\tau_{\varphi} \equiv r^{-1} \partial_{\varphi}  T$, the Poisson equation \eref{simpleT} is written as,
\begin{align}
\nabla \cdot \nabla T &=
\frac{1}{r} \frac{\partial}{\partial r} \left( r \tau_r \right)
+ \frac{1}{r} \frac{\partial}{\partial \varphi} \tau_\varphi
= -\frac{\rho}{\kappa} J^2.
\label{eq25}
\end{align}
Since we find that $\tau_r \ll \tau_\varphi$ in the vicinity of the corner $r=0$ in our numerical result, we can neglect the first term in \eref{eq25}.
Thus, we can express the gradient $\tau_{\varphi}$ at the isothermal boundary ($\varphi = \pi /2 $) with a small $r$ by integrating \eref{eq25} as,
\begin{align}
\tau_{\varphi} \left( \frac{\pi}{2}, r \right) =
\left. \tau_{\varphi} \right|_{\varphi=0} - \frac{\rho}{\kappa} \int_0^{\pi / 2} r J^2 d \varphi.
\end{align}
Since $J$ is nearly isotropic, we can replace $J$ in the integral by $J_x$ at the bottom boundary, $J |_{\varphi=0} = -J_x |_{y=0} \equiv J(r)$ (note that $J_x <0$ and $J_y=0$ at the bottom boundary).
With this approximation, we obtain
\begin{align}
\tau_{\varphi} \left( \frac{\pi}{2}, r \right) &=
\left. \tau_{\varphi} \right|_{\varphi=0} - \frac{\rho}{\kappa} \frac{\pi}{2} r J(r)^2.
\label{delyT1}
\end{align}
Noting that $M$ is not essential for the cooling effect (see \sref{cooling mechanism}), we assume $M=0$ in \eref{simpleeqsBCT} for simplicity to obtain
\begin{align}
\left. \tau_{\varphi} \right|_{\varphi = 0} = \left. \partial_y T \right|_{y = 0} = -\frac{NTB}{\kappa} J(r).
\label{delyT2}
\end{align}
Using \eref{delyT2} in \eref{delyT1}, we have
\begin{align}
\tau_{\varphi} \left( \frac{\pi}{2}, r \right) =
-\frac{NTB}{\kappa} J(r) - \frac{\rho}{\kappa} \frac{\pi}{2} r J(r)^2 .
\end{align}
As illustrated in \fref{fig7},
temperatures lower than $T_\mathrm{left}=T(\varphi=\pi/2)$ emerge if $\tau_{\varphi} (\pi/2, r) > 0$, namely if
\begin{align}
B>
\frac{\pi}{2} \frac{\rho}{(-N) T} r J(r)
= \frac{\pi}{2} \frac{E_0 \cos  \theta_{\mathrm{H}}}{(-N) T} r e^{\gamma (r)},
\label{sec4-49}
\end{align}
with $\gamma(r) \equiv \gamma(\xi=r/L_x,\eta=0)$, and we used $N<0$ in the derivation.
As will be shown below, the right-hand side tends to $0$ with $r \rightarrow 0$, although the current $J(r)$ diverges with $r \rightarrow 0$ (see \fref{fig4}).
On the bottom boundary, we have from \eref{gamma},
\begin{align}
\gamma (r)
 &= 4 \theta_{\mathrm{H}}
\left(
\frac{1}{2 \pi}
\ln \left[ \cot \left( \frac{\pi r}{2 L_x} \right) \right]
\right. \notag \\
&
\left.
- \sum_{n=1}^{\infty}
\frac{\cos \left[ (2n -1) \pi \frac{r}{L_x} \right] }{(2n-1) \pi}
\left\{
1- \tanh \left[ (2n-1) \frac{\lambda \pi}{2} \right]
\right\}
\right).
\end{align}
The second term in the large round brackets is less than $0$ for $r  \rightarrow 0$, and thus from \eref{EQ23},
\begin{align}
J(r) &< E_0 \frac{\cos \theta_{\mathrm{H}}}{\rho} 
\exp \left\{ \frac{2 \theta_{\mathrm{H}}}{\pi} \ln \left[ \cot \left( \frac{\pi r}{2 L_x} \right) \right] \right\}
\notag \\
 &= E_0 \frac{\cos \theta_{\mathrm{H}}}{\rho} 
\left[ \left( \frac{\pi r}{2 L_x} \right)^{ - \frac{2 \theta_{\mathrm{H}}}{\pi}} + O(r)  \right].
\end{align}
Since $0 < 2 \theta_{\mathrm{H}} / \pi < 1$, we have $r J(r) \rightarrow 0$ for $r \rightarrow 0$.

With a fixed $B$, an area within the distance $r$ from the left-bottom corner  becomes colder than the isothermal boundary ($\varphi = \pi / 2$) for $r$ satisfying \eref{sec4-49}.
Alternatively, for fixed $r$ ($\sim 0$), \eref{sec4-49} gives a threshold magnetic field for the position $r$ to become colder than the isothermal boundary, (which can be made, in principle, arbitrarily small by letting $r \rightarrow 0$).
Practical thresholds for the numerical calculation is given by the mesh spacing $d$.
By setting $r=d=0.1\,\,\mathrm{\mu m}$, we have a threshold magnetic field $0.03\,\,\mathrm{T}$, which is consistent with our numerical results for $B=0.03$ T (not shown).

\section{Discussion}
\label{sec_discussion}
We have shown that the electron temperature can be cooled down by a dc electric current with an arbitrarily small magnetic field. The emergence of the cooled area is qualitatively consistent with the cooling inferred by the experimentally observed sign reversal of the Nernst signal~\cite{Fujita} described in the Introduction. Since the present calculation is performed on much simplified setups compared to the actual experimental arrangements, however, it is rather difficult to unambiguously relate the present results to the experimental findings at this stage. In this section, we point out noticeable discrepancies between the setups in our calculation and the experiment, and discuss modifications necessary to be made in the theoretical treatment to fill the gap and to facilitate more precise and quantitative comparison in the future study.

First, the calculated temperature decrement is extremely small, order of 10$^{-2}$ $\mu$K\@. It is rather unlikely that the effect caused by such a small temperature change can be experimentally detected. It will be necessary to alter our boundary conditions to reflect the experimental conditions more precisely. Above all, we set the temperature of boundaries immediately to the left and right of the ``heater section'' fixed, while in the experiment the temperature is fixed at the left and right contact pads (hatched rectangles in \fref{fig1}(a)) separated $\sim$300 $\mu$m away from the heater section (the region below the front gate depicted by the gray rectangle in \fref{fig1}(a)). It seems plausible to expect from the mechanism described in \sref{chap3} that the separation allows larger variation of the temperature at the left-bottom corner of the heater section.

Our boundary condition also neglects the possible effects brought about by the heating current $I_\mathrm{h}$ passing through the regions with different carrier densities (regions with and without the front gate, see \fref{fig1}(a)) hence having differing transport coefficients. In fact, it has been pointed out \cite{Nakagawa05,Nagai07} that the heat can be either absorbed or emitted by $I_\mathrm{h}$ at the boundary between the regions having different filling factors, owing to the difference in the Peltier constant in the two regions, resulting in cooled or heated areas. This can be an alternative candidate for the origin of the cooling effect distinct from the mechanism discussed in the present paper.

The fixed potential difference between the right and left boundaries $\Delta \phi$ is also at variance with the experiment, in which the total current $J_\mathrm{tot}$ is fixed. According to equations \eref{Hall} and \eref{JtotDphi}, $J_\mathrm{tot}$ decreases with increasing $B$ if $\Delta \phi$ is kept constant. We can envisage larger temperature change if we increase $\Delta \phi$ with $B$ to keep $J_\mathrm{tot}$ constant as in the experiment. (We found difficulty, however, in the convergence of the numerical calculation if we set a larger value for $\Delta \phi$.) 

So far, our calculation is limited to rather small magnetic field ($B\leq 0.2$ T) for technical reasons, especially for the bad convergence of the calculation. Noting that both the temperature decrement and the area of the cooled region increase with increasing magnetic field (see figure 3), we can expect much more pronounced cooling effect if we can extend our calculation to larger magnetic field.  

Second, the sign reversal (corresponding to the appearance of the cooled area) was observed only when the magnetic field was large enough ($B>1.8$ T) and only for the value of $V_\mathrm{g}$ at which electrons occupy less than half of the topmost Landau level (see figure 1 (c)). This can be related to the magnitude and the sign of the Ettingshausen coefficient $NT$, or equivalently, to those of the Nernst coefficient $N$. (Note that the Ettingshausen and Nernst coefficients are related by the Kelvin-Onsager relation.) As mentioned earlier, we used, in the present calculation, the value of $N(<0)$ at $B$=0 for simplicity, neglecting the $B$-dependence. In reality, the magnitude $|N|$ decreases with increasing $B$, until Landau levels are clearly resolved. In a quantizing magnetic field, $|N|$ again takes a large value when the Fermi energy $E_F$ lies in a (disorder broadened) Landau level, with the sign of $N$ alternating depending on whether $E_F$ is below or above the center of the Landau level, namely, whether the energy derivative of the density of states is positive (electron-like) or negative (hole-like)~\cite{Fletcher1}\footnote{Note that $N$ in the present paper corresponds to $S_{xy}/B$ in \cite{Fletcher1}.}. It can readily be seen from the discussion in \sref{chap3} that the cold and hot areas appearing in the left-bottom and right-top corners, respectively, interchange their roles when the sign of $N$ is inverted. The conditions for $B$ and $V_g$ mentioned above thus can be interpreted as the condition that $N$ possesses large enough magnitude (for the Ettingshausen effect to have sufficient strength) and the appropriate sign (so that the cooled area is generated on the side adjacent to the main Hall bar), respectively.

Third, in a 2DEG in the transition region between two adjacent quantum Hall states, the current is carried by both the bulk extended state of the topmost Landau level and the edge states from lower Landau levels. Although the edge states can have significant impact on the distributions of the electric and heat currents \cite{Granger09}, and hence on the temperature distribution, our calculation amounts to neglecting the edge states altogether.

Apparently, much improvement has to be made to quantitatively explain the experiment motivated our study. We believe, however, that our simplified approach has been advantageous to pinpoint the very essence of the cooling mechanism.

\section{Conclusions}
\label{chap4}

We have investigated the temperature distribution induced by a dc current in a 2DEG subjected to a perpendicular magnetic field, surrounded by isopotential-isothermal (left and right) and insulated-adiabatic (top and bottom) boundaries (\fref{fig1}(b), \fref{fig2}). By numerically solving the nonlinear Poisson equations \eref{PHIpoisson1} and \eref{Tpoisson1}, we have demonstrated that an area having the temperature lower than the isothermal boundaries (kept at $40\,\,\mathrm{mK}$) appears in the vicinity of one of the corners where the electric current density is highly concentrated. The cooling is ascribed to the Ettinghausen effect, which  pumps the heat away from the adiabatic boundary. The adiabatic condition \eref{eqsBCT} at the boundary requires the temperature gradient to be generated, with the resulting thermal diffusion canceling out the thermal current due to the Ettingshausen effect. We have shown that, owing to the temperature gradient, the cooled area emerges with an arbitrarily small magnetic field, although the area shrinks within closer proximity of the corner with decreasing magnetic field. 

Although the present study is motivated by the recent experiment \cite{Fujita}, the relation of the present mechanism to the experiment remains rather unclear. Nevertheless, the confirmation of the presence of the rather counterintuitive current-induced cooling effect, as well as the identification of the mechanism responsible for the effect, appears to be important in its own right, and underlines the complication brought about by the thermoelectric and thermomagnetic effects.

\ack
The present authors are grateful to Prof.\ A.\ Kamitani for valuable comments and are indebted to Prof.\ H.\ Okumura for his advice on the derivation fo the nonlinear Poisson equations.
The work is supported by Grant-in-Aid for Scientific Research (B) No.\ 20340101 and (C) No.\ 22560004 from Japan Society for the Promotion of Science, as well as by Research Fund of Tokyo University of Science, by the Izumi Science and Technology Foundation, by NINS program for cross-disciplinary study (NIFS10KEIN0160) and by the National Institutes of Natural Sciences undertaking Forming Bases for Interdisciplinary and International Research through Cooperation Across Fields of Study and Collaborative Research Program 
No.\ NIFS08KEIN0091.

\appendix
\section{\label{appendix1}Derivation of the nonlinear Poisson equations \eref{PHIpoisson1} and \eref{Tpoisson1}}

In this Appendix, we present, for completeness, the derivation of the nonlinear Poisson equations \eref{PHIpoisson1} and \eref{Tpoisson1} from the transport equations \eref{EC potential} and \eref{JQ} in the steady-state. We basically follow the prescription presented in \cite{Okumura1} and \cite{Okumura3}. See also \cite{Harman2}.

In the steady state ($\partial / \partial t =0$), we have, from the continuity equations, 
\begin{align}
&\nabla \cdot \boldsymbol{J}
= 0,
\label{AP1} 
\end{align}
and
\begin{align}
&\nabla \cdot \boldsymbol{J_U}
=\nabla \cdot ( \boldsymbol{J_Q} + \phi \boldsymbol{J}) = 0.
\label{AP2} 
\end{align}

Taking the divergence of \eref{EC potential}, we have, with the aid of \eref{AP1}:
\begin{align}
- \nabla^2 \phi
=&
\nabla \cdot \left[
\rho \boldsymbol{J}
+ R \boldsymbol{B} \times \boldsymbol{J} 
+ \alpha \nabla T 
+ N \boldsymbol{B} \times \nabla T
\right]
\notag \\
=& 
\nabla \rho \cdot \boldsymbol{J} 
+ \nabla R \cdot (\boldsymbol{B} \times \boldsymbol{J} )
+  R [ \nabla \cdot (\boldsymbol{B} \times \boldsymbol{J} )]
\notag \\
&
+\nabla \alpha \cdot \nabla T 
+ \alpha \nabla^2 T 
\notag \\
&+ \nabla N \cdot (\boldsymbol{B} \times \nabla T)
+  N [ \nabla \cdot (\boldsymbol{B} \times \nabla T)].
\label{newAP3}
\end{align}
Noting that, under the temperature gradient $\nabla T$, an arbitrary transport coefficient $X$ depends on $(x, y)$ through its temperature dependence, the gradient of $X$ can be expressed as
\begin{align}
\nabla X = \frac{d X}{d T} \nabla T.
\label{AP4} 
\end{align}
We thus can rewrite \eref{newAP3} as follows:
\begin{align}
- \nabla^2 \phi
=& 
\frac{d \rho}{d T} (\boldsymbol{J} \cdot \nabla T ) 
\notag \\
&
+ \frac{d R}{d T} [ \nabla T \cdot (\boldsymbol{B} \times \boldsymbol{J} )]
+  R [ \nabla \cdot (\boldsymbol{B} \times \boldsymbol{J} )]
\notag \\
&
+\frac{d \alpha}{d T} ( \nabla T )^2 
+ \alpha \nabla^2 T 
\notag \\
&+\frac{d N}{d T} [ \nabla T \cdot (\boldsymbol{B} \times \nabla T)]
+  N[ \nabla \cdot  (\boldsymbol{B} \times \nabla T)].
\label{newAP5}
\end{align}
Using formulas for vector operation, \eref{newAP5} becomes
\begin{align}
- \nabla^2 \phi
=& 
\frac{d \rho}{d T} (\boldsymbol{J} \cdot \nabla T ) 
+ \frac{d R}{d T} [ \nabla T \cdot (\boldsymbol{B} \times \boldsymbol{J} )]
\notag \\
&
+  R [\boldsymbol{J} \cdot (\nabla \times \boldsymbol{B}) - \boldsymbol{B} \cdot ( \nabla \times \boldsymbol{J} )]
\notag \\
&
+\frac{d \alpha}{d T} ( \nabla T )^2 
+ \alpha \nabla^2 T 
\notag \\
&+  N \nabla T \cdot  (\nabla \times \boldsymbol{B}).
\label{newAP6}
\end{align}
Using further the Maxwell equation $\nabla \times \boldsymbol{B} = \mu \boldsymbol{J} + \epsilon \mu (d \boldsymbol{E} / d t)= 0$ in the steady state, omitting the negligibly small term $\mu J$, \eref{newAP6} becomes
\begin{align}
- \nabla^2 \phi
=& 
\frac{d \rho}{d T} (\boldsymbol{J} \cdot \nabla T ) 
\notag \\
&+ \frac{d R}{d T} [ \nabla T \cdot (\boldsymbol{B} \times \boldsymbol{J} )]
-  R [ \boldsymbol{B} \cdot ( \nabla \times \boldsymbol{J} ) ].
\notag \\
&+\frac{d \alpha}{d T} ( \nabla T )^2 
+ \alpha \nabla^2 T .
\label{AP26}
\end{align}

Here we calculate $\nabla \times \boldsymbol{J}$ from \eref{EC potential} under the conditions \eref{AP1} and \eref{AP2}.
The rotation of \eref{EC potential} gives
\begin{align}
0
&=- \nabla \times \nabla \phi
\notag \\
&= \nabla \times 
[\rho \boldsymbol{J} 
+ R \boldsymbol{B} \times \boldsymbol{J} 
+ \alpha \nabla T 
+ N \boldsymbol{B} \times \nabla T]
\notag 
\\
&= 
\frac{d \rho}{d T} ( \nabla  T \times \boldsymbol{J})
+ \rho ( \nabla \times \boldsymbol{J})
\notag \\
& \hspace{1pc}
+
(\boldsymbol{J} \cdot \nabla)( R \boldsymbol{B})
-(R \boldsymbol{B} \cdot \nabla ) \boldsymbol{J}
- \boldsymbol{J} [ \nabla \cdot (R \boldsymbol{B} )]
\notag \\
& \hspace{1pc}
+R \boldsymbol{B} (\nabla \cdot \boldsymbol{J})
\notag \\
& \hspace{1pc}
+ \frac{d \alpha}{d T} (\nabla \times \nabla T )
+ \alpha  ( \nabla \times \nabla T)
\notag \\
& \hspace{1pc}
+
(\nabla T \cdot \nabla) ( N \boldsymbol{B})
- ( N \boldsymbol{B}  \cdot \nabla ) \nabla T 
\notag \\
& \hspace{1pc}
- \nabla T [ (\nabla \cdot (N \boldsymbol{B})]
+ N \boldsymbol{B} (\nabla \cdot \nabla T)
\notag \\
&= 
\frac{d \rho}{d T} ( \nabla  T \times \boldsymbol{J})
+ \rho ( \nabla \times \boldsymbol{J})
\notag \\
& \hspace{1pc}
+
 \frac{d R}{d T}  \left( \boldsymbol{J} \cdot \nabla T  \right) \boldsymbol{B}
+
\left[
N \nabla^2 T
+ \frac{d N}{d T} (\nabla T)^2
\right] \boldsymbol{B}.
\end{align}
We thus obtain
\begin{align}
 \nabla \times \boldsymbol{J}
&=
- \frac{1}{\rho}
\left\{
\frac{d \rho}{d T} ( \nabla  T \times \boldsymbol{J})
 \right.
\notag \\
&
\left.
+
\left[
 \frac{d R}{d T}  \left( \boldsymbol{J} \cdot \nabla T  \right) 
+
N \nabla^2 T
+ \frac{d N}{d T} (\nabla T)^2
\right] \boldsymbol{B}
\right\}.
\label{AP20}
\end{align}

Substituting \eref{AP20} into \eref{AP26}, we obtain the following equation:
\begin{align}
- \nabla^2 \phi
=& 
\frac{d \rho}{d T} (\boldsymbol{J} \cdot \nabla T ) 
\notag \\
&+ \frac{d R}{d T} [ \nabla T \cdot (\boldsymbol{B} \times \boldsymbol{J} )]
+ \frac{R}{\rho}
\frac{d \rho}{d T} 
[ \boldsymbol{B} \cdot ( \nabla  T \times \boldsymbol{J}) ]
\notag \\
&+ \frac{R B^2}{\rho}
\left[
 \frac{d R}{d T}  \left( \boldsymbol{J} \cdot \nabla T  \right) 
 +
N \nabla^2 T
+ \frac{d N}{d T} (\nabla T)^2
\right]
\notag \\
&+\frac{d \alpha}{d T} ( \nabla T )^2 
+ \alpha \nabla^2 T
\notag
\\
=& 
\left(
\frac{d \alpha }{d T} + \frac{R B^2}{\rho } \frac{d N }{d T}
\right)
(\nabla T)^2
\notag \\
\hspace{1pc}
&
-
\left(
\frac{R}{\rho} \frac{d \rho}{d T} - \frac{d R}{d T}
\right)
[\nabla T \cdot (\boldsymbol{B} \times \boldsymbol{J})]
\notag \\
\hspace{1pc}
&
+
\left(
\frac{d \rho}{d T} + \frac{R B^2}{\rho} \frac{d R}{d T}
 \right)
(\boldsymbol{J} \cdot \nabla T)
\notag \\
\hspace{1pc}
&+
\left(
\alpha + \frac{R N B^2}{ \rho} 
\right)
\nabla^2 T.
\label{AP28}
\end{align}
By substituting \eref{Tpoisson1} (to be derived below) in \eref{AP28}, we arrive at equation \eref{PHIpoisson1}.

Next, we derive the Poisson equation for $T$, \eref{Tpoisson1}.
We obtain the following equation from the divergence of $\boldsymbol{J_U}$ using equations \eref{AP1} and \eref{AP2}:
\begin{align}
&\nabla \cdot (\phi \boldsymbol{J}) 
+ \nabla \cdot  (\alpha T \boldsymbol{J})  
+ \nabla \cdot  [N T (\boldsymbol{B} \times \boldsymbol{J} )]
\notag \\
&\hspace{1pc}
- \nabla \cdot  (\kappa \nabla T )
+ \nabla \cdot  [\kappa M (\boldsymbol{B} \times \nabla T)]
\notag \\
&=
\nabla \phi  \cdot \boldsymbol{J}
+ \nabla  \alpha  \cdot  (T \boldsymbol{J}) 
+ \alpha \nabla T  \cdot  \boldsymbol{J}
\notag \\
&\hspace{1pc}
+ (T \nabla N + N \nabla T)  \cdot (\boldsymbol{B} \times \boldsymbol{J}) 
+NT [ \nabla \cdot (\boldsymbol{B} \times \boldsymbol{J})]
\notag \\
&\hspace{1pc} 
- \nabla \kappa \cdot \nabla T 
-  \kappa \nabla^2 T 
+(M \nabla \kappa + \kappa \nabla M) \cdot (\boldsymbol{B} \times \nabla T)
\notag \\
&\hspace{1pc}
+ \kappa M [ \nabla \cdot (\boldsymbol{B} \times \nabla T)]
= 0.
\label{AP3}
\end{align}
Using \eref{AP4}, we have
\begin{align}
&\nabla \phi  \cdot \boldsymbol{J}
+ \left( T \frac{d \alpha}{d T} + \alpha \right)
( \nabla T  \cdot  \boldsymbol{J})
\notag \\
&\hspace{1pc}
+ \left( T \frac{d N}{d T} + N \right) 
[ \nabla T \cdot (\boldsymbol{B} \times \boldsymbol{J}) ]
+NT [ \nabla \cdot (\boldsymbol{B} \times \boldsymbol{J})]
\notag \\
&\hspace{1pc} 
- \frac{d \kappa}{d T}  ( \nabla T)^2 
-  \kappa \nabla^2 T 
\notag \\
&\hspace{1pc} 
+ \left(M \frac{ d \kappa}{d T} + \kappa \nabla \frac{d M}{d T} \right)
[ \nabla T \cdot (\boldsymbol{B} \times \nabla T)]
\notag \\
&\hspace{1pc} 
+ \kappa M [ \nabla \cdot (\boldsymbol{B} \times \nabla T)]
=0.
\label{AP5}
\end{align}
Using again formulas for vector operation, we rewrite the above equation as,
\begin{align}
&\nabla \phi  \cdot \boldsymbol{J}
+ \left( T \frac{d \alpha}{d T} + \alpha \right)
( \nabla T  \cdot  \boldsymbol{J})
\notag \\
&\hspace{1pc}
+ \left( T \frac{d N}{d T} + N \right) 
[ \nabla T \cdot (\boldsymbol{B} \times \boldsymbol{J}) ]
\notag \\
&\hspace{1pc}
+NT [
\boldsymbol{J} \cdot (\nabla \times \boldsymbol{B})
- \boldsymbol{B} \cdot (\nabla \times \boldsymbol{J})]
\notag \\
&\hspace{1pc} 
- \frac{d \kappa}{d T}  ( \nabla T)^2 
-  \kappa \nabla^2 T 
\notag \\
&\hspace{1pc} 
+ \kappa M 
[ \nabla T \cdot (\nabla \times \boldsymbol{B})]
=0,
\label{AP14}
\end{align}
which reduces, with the Maxwell equation $\nabla \times B=0$, to
\begin{align}
\kappa \nabla^2 T 
=
&\nabla \phi  \cdot \boldsymbol{J}
+ \left( T \frac{d \alpha}{d T} + \alpha \right)
( \nabla T  \cdot  \boldsymbol{J})
\notag \\
&
+ \left( T \frac{d N}{d T} + N \right) 
[ \nabla T \cdot (\boldsymbol{B} \times \boldsymbol{J}) ]
\notag \\
&
-NT [\boldsymbol{B} \cdot (\nabla \times \boldsymbol{J})]
- \frac{d \kappa}{d T}  ( \nabla T)^2. 
\end{align}

We substitute \eref{EC potential} into the above equation and obtain
\begin{align}
\kappa \nabla^2 T 
=&-[
\rho \boldsymbol{J} 
+ R ( \boldsymbol{B} \times \boldsymbol{J} )
+ \alpha \nabla T 
+ N (\boldsymbol{B} \times \nabla T)] \cdot \boldsymbol{J}
\notag \\
&
+ \left( T \frac{d \alpha}{d T} + \alpha \right)
( \nabla T  \cdot  \boldsymbol{J})
\notag \\
&
+ \left( T \frac{d N}{d T} + N \right) 
[ \nabla T \cdot (\boldsymbol{B} \times \boldsymbol{J}) ]
\notag \\
&
-NT [\boldsymbol{B} \cdot (\nabla \times \boldsymbol{J})]
- \frac{d \kappa }{d T} (\nabla T)^2
\notag
\\
=&
- \rho J^2
+ T \frac{d \alpha}{d T}  ( \nabla T  \cdot  \boldsymbol{J})
\notag \\
&+ \left( T \frac{d N}{d T} + 2 N \right) 
[ \nabla T \cdot (\boldsymbol{B} \times \boldsymbol{J}) ]
\notag \\
&-NT \left[
\boldsymbol{B} \cdot (\nabla \times \boldsymbol{J})
\right]
- \frac{d \kappa }{d T} (\nabla T)^2.
\label{AP10}
\end{align}
By substituting \eref{AP20} into \eref{AP10}, we arrive at \eref{Tpoisson1}.

\bibliography{refAE}

\end{document}